# Robust energy selective tunneling readout of singlet triplet qubits under large magnetic field gradient

Wonjin Jang[1†], Jehyun Kim[1†], Min-Kyun Cho[1], Hwanchul Chung[2], Sanghyeok Park[1], Jaeun Eom[1], Vladimir Umansky[3], Yunchul Chung[2], and Dohun Kim[1]*

[1]Department of Physics and Astronomy, and Institute of Applied Physics, Seoul National University, Seoul 08826, Korea

[2] Department of Physics, Pusan National University, Busan 46241, Korea

[3]Braun Center for Submicron Research, Department of Condensed Matter Physics, Weizmann Institute of Science, Rehovot 76100, Israel

[†]These authors contributed equally to this work

*Corresponding author: dohunkim@snu.ac.kr

**Abstract**

Fast and high-fidelity quantum state detection is essential for building robust spin-based quantum information processing platforms in semiconductors. The Pauli spin blockade (PSB)-based spin-to-charge conversion and its variants are widely used for the spin state discrimination of two-electron singlet–triplet ($ST_0$) qubits; however, the single-shot measurement fidelity is limited by either the low signal contrast, or the short lifetime of the triplet state at the PSB energy detuning, especially due to strong mixing with singlet states at large magnetic field gradients. Ultimately, the limited single-shot measurement fidelity leads to low visibility of quantum operations. Here, we demonstrate an alternative method to achieve spin-to-charge conversion of $ST_0$ qubit states using energy selective tunneling between doubly occupied quantum dots (QDs) and electron reservoirs. We demonstrate a single-shot measurement fidelity of 90% and an S–$T_0$ oscillation visibility of 81% at a field gradient of 100 mT ($\sim 500\,MHz \cdot h \cdot (g* \cdot \mu_B)^{-1}$); this allows single-shot readout with full electron charge signal contrast and, at the same time, long and tunable measurement time with negligible effect of relaxation even at strong magnetic field gradients. Using an rf-sensor positioned opposite to the QD array, we apply this method to two $ST_0$ qubits and show high-visibility readout of two individual single-qubit gate operations is possible with a single rf single-electron transistor sensor. We expect our measurement scheme for two-electron spin states can be applied to various hosting materials and provides a simplified and complementary route for multiple qubit state detection with high accuracy in QD-based quantum computing platforms.

**Introduction**

The assessment of general quantum information processing performance can be divided into that of state initialization, manipulation, and measurement. Rapid progress has been made in semiconductor quantum dot (QD) platforms, with independent demonstrations of, for example, high-fidelity state initialization of single and double QD spin qubits[1–3], high-fidelity quantum control with resonant microwaves[4–8] and non-adiabatic pulses[1,9,10], and high-fidelity state measurements using spin-to-charge conversion[3,11–19]. However, the high visibility of a quantum operation requires high fidelity in all stages of the quantum algorithm execution, which has been demonstrated in only a few types of spin qubits so far[4,6,7,10,20,21].

For double QD two-electron spin qubits, the Pauli spin blockade (PSB) phenomenon is typically used for discriminating spin-singlet (S) and -triplet ($T_0$) states where different spin states are mapped according to the difference in the relative charge occupation of two electrons inside the double QD, which is detected by a nearby electrometer[22–24]. As the spin-dependent signal deterministically appear at the measurement phase defined by the pulse sequence at the PSB, the measurement window can be shortened to the limit which allows enough signal to noise ratio (SNR) to discriminate the different spin signal[13], and such can lead to high measurement bandwidth. However, depending on the device design, the signal contrast can be small compared to the signal of one electron, especially when the charge sensor position in the device is not aligned with the QD axis. This issue is particularly problematic in recent multiple QD designs[25–29], where the charge sensor positioned opposite to the qubit array increases the range of QDs detectable by one sensor, but renders sensitive measurement of the relative electron position between nearest-neighbor QDs difficult.

Moreover, the spatial magnetic field difference $\Delta B_{//} = |B_{\mathrm{L}//} - B_{\mathrm{R}//}|$, where the $B_{\mathrm{L}//}$ ($B_{\mathrm{R}//}$) denotes the magnetic field strength parallel to the spin quantization axis at the left (right) dot, provides relaxation pathways through (1,1)T$_0$–(1,1)S mixing and rapid (1,1)S to (2,0)S tunneling in the PSB region as shown in the solid green regions in Fig. 1a, and normal PSB readout is difficult under large $\Delta B_{//}$. For example when $\Delta B_{//} > 200\ MHz \cdot h \cdot (g* \cdot \mu_{\mathrm{B}})^{-1}$, where $h$ is the Planck's constant, $g*$ is the electron g-factor in GaAs, and $\mu_{\mathrm{B}}$ is the Bohr magneton, the fast spin relaxation is known to lead to vanishing oscillation visibility[30]. As most QD spin qubit platforms utilize sizeable intrinsic[2,31,32] or extrinsic[33] $\Delta B_{//}$ to realize individual qubit addressing and high-fidelity single- and two-qubit operations[4,6,34,35], it is important to develop fast readout techniques that enable high-fidelity spin detection even at large $\Delta B_{//}$. So far, visibility higher than 95% using PSB readout can be achieved only for small $\Delta B_{//}$ despite the method's high measurement bandwidth[13,18].

These limitations of conventional PSB readout have been addressed in previous works, and several variants of the PSB readout have been developed for various QD systems[14–17]. In the latched readout scheme[14], the lack of the reservoir on one side of the double QD enables spin conversion to the (1,0) or (2,1) charge state, enhancing the signal contrast. In Ref [15], singlet–triplet ($ST_0$) qubit readout was performed in a triple QD to isolate the middle QD from the reservoirs, and the qubit state conversion to a metastable charge state enabled robust, high-fidelity qubit readout. While these techniques enhance the signal contrast to the full electron charge, the explicit demonstration of such methods combined with high-fidelity operation under large $\Delta B_{//}$ ($> 200$ $MHz \cdot h \cdot (g * \cdot \mu_B)^{-1}$) has not been reported to date. We stress that it is unclear whether the readout near the (2,1) charge transition[15,17] will not suffer from the fast $T_0$ relaxation if the spin mixing rate due to $\Delta B_{//}$ is comparable to the (1,1)$T_0$ − (2,1) tunneling rate. We note here that unlike the readout methods near the (2,1) charge transition[15,17], $T_0$ relaxation pathway is inherently absent at the readout position of this work, as both the S and $T_0$ state occupy the (2,0) charge state as we describe below in detail. On the other hand, Orona, L. A. et al.[16] reported the shelving readout technique, whereby one of the qubit states is first converted to the $T_+$ state through fast electron exchange with the reservoir to prevent mixing with the (1,1)S state, enabling high-visibility readout of the $ST_0$ spin qubit. They showed explicitly that single-shot readout is possible even for $\Delta B_{//}$ ~ 180 mT (~900 $MHz \cdot h \cdot (g * \cdot \mu_B)^{-1}$) by optimizing the shelving pulse sequence. However, the technique relies on PSB for final spin-to-charge conversion and is expected to be effective only when the charge sensor is sensitive to the relative position of electrons in the double QD.

Here, we demonstrate the energy selective tunneling (EST) readout, commonly called

Elzerman readout[11], of $ST_0$ qubits under large $\Delta B_{//}$, accomplishing both signal enhancement, due to one electron tunneling, and long measurement time, enabling a robust single-shot readout. Unlike previous works, which demonstrated independent enhancement of the signal contrast and measurement time through intermediate spin or charge state conversion steps, our scheme does not require additional state conversion during the readout. Using large voltage modulation by rapid pulsing with $\varepsilon$ ranging from the PSB-lifted (2,0) to the deep (1,1) charge regions, where the exchange coupling $J(\varepsilon)$ is turned off, we explicitly demonstrate a single-shot measurement fidelity of $90 \pm 1.3$ % and an S–$T_0$ oscillation visibility of 81% at $\Delta B_{//} \sim$ 100 mT, corresponding to an oscillation frequency of 500 MHz. Furthermore, we demonstrate the detection of coherent operation of two individual $ST_0$ qubits in a quadruple QD array with a single rf-reflectometry line. We stress that we combine previous methods which individually demonstrated the Elzerman readout of the two electron spin states[12], large $\Delta B_{//}$ generation with micromagnet[33], high fidelity control of the $ST_0$ qubit [36], and robust measurement within a single quantum processor yielding a record high quantum oscillation visibility in large $\Delta B_{//}$. We also note that this is achieved at the expense of high bandwidth of PSB readout due to EST readout's intrinsic timing uncertainty in tunneling events. However the achieved measurement time on the order of 100 $\mu s$ in this work using EST readout is still useful for future application to fast spin state readout, for example single-shot readout-based Bayesian estimation[37]. In this paper, we describe the proposed EST readout method in detail, compare it with the conventional PSB readout, and suggest possible routes for its further optimization.

**Results**

**Energy selective tunneling readout**

The blue rectangular regions in Fig. 1a show the position of $\varepsilon$ and the energy level configuration used for EST state initialization and readout. At this readout point, the PSB is lifted, and both S and $T_0$ levels can first occupy the (2,0) charge state, the energies of which are separated by $ST_0$ splitting typically in the order of ~ 25–30 GHz [38], depending on the dot-confining potential. Near the (1,0) - (2,0) electron transition, the electrochemical potential of the reservoir resides between these states, which enables the EST of the $ST_0$ qubits. As discussed in detail below, we observe the single-shot spin-dependent tunneling signal where one electron occupying an excited orbital state of the (2,0)$T_0$ state tunnels to the reservoir to form the (1,0) charge state, leading to an abrupt change in the sensor signal, and predominantly initializes back to the energetically favorable (2,0)S state. In contrast, no tunneling occurs for the (2,0)S state (see Fig. 1a, blue right panel).

We study a quadruple QD array with an rf single-electron transistor (rf-set) sensor consisting of Au/Ti metal gates on top of a GaAs/AlGaAs heterostructure, where a 2D electron gas (2DEG) is formed approximately 70 nm below the surface (Fig. 1b). A 250 nm-thick rectangular Co micromagnet with large shape anisotropy was deposited on top of the heterostructure to generate stable $\Delta B_{//}$ for $ST_0$ qubit operation[33,36,39,40] (see methods section for fabrication details). The device was placed on a plate in a dilution refrigerator at ~20 mK and an in-plane magnetic field $B_{z,ext}$ of 225 mT was applied. To demonstrate the EST readout in the experiment, we independently operated and readout two $ST_0$ qubits ($Q_L$ and $Q_R$) in the non-interacting regime by blocking $Q_L$–$Q_R$ tunneling using appropriate gate voltages. We monitored the rf-reflectance of the rf-set sensor (Fig. 1b, yellow dot) for fast single-shot charge occupancy detection in the $\mu s$ time scale[41,42]. The intra qubit tunnel couplings for both $Q_L$ and $Q_L$ and $Q_R$ were tuned above 8 GHz to suppress unwanted Landau–Zener–Stuckelberg interference under fast $\varepsilon$ modulation, and we estimated the electron temperature to be approximately 230 mK (see also Supplementary Note 1).

We first locate appropriate EST readout points in the charge stability diagrams. Figure 1c (1d) shows the relevant region in the stability diagram for the $Q_L$ ($Q_R$) qubit operation as a function of two gate voltages $V_1$ ($V_3$) and $V_2$ ($V_4$). We superimpose the cyclic voltage pulse, sequentially reaching I – W – O – W – R points in the stability diagram (see Fig. 1c and 1d) with a pulse rise time of 200 ps. During the transition from the point W to point O stage, the pulse brings the initialized (2,0)S state to the deep (1,1) region non-adiabatically, and the time evolution at point O results in coherent S-$T_0$ mixing due to $\Delta B_{//}$. The resultant non-zero $T_0$ probability is detected at the I/R point. For this initial measurement, the duration of each pulse stage was not strictly calibrated, but the repetition rate was set to 10 kHz. The resulting 'mouse-bite' pattern inside the (2,0) charge region (Fig. 1c., boundary marked by the red dashed line) implies the (1,0) charge occupancy within the measurement window, which arises from the EST of the $ST_0$ qubit states averaged over $100 \, \mu s$. For comparison, we note that the PSB readout signal with a similar pulse sequence is not clearly visible in the main panel of Fig. 1c in the time-averaged manner due to fast relaxation, as described above. The inset in Fig. 1c shows the PSB readout signal measured by gated (boxcar) integration (see Supplementary Note 2), where an approximately 100 ns gate window was applied immediately after the pulse sequence. This difference in the available range of measurement time scale clearly contrasts two distinct readout mechanisms for the spin-to-charge conversion of $ST_0$ qubits.

The PSB and EST readouts are systematically compared through time-resolved relaxation measurements, which also serve as calibration of the readout parameters for EST readout visibility optimization. Fig. 2a (2b) shows the relaxation of the sensor signal as a function of waiting time $\tau$ before reaching the measurement stage, using the pulse sequence shown in the inset of Fig. 2a (2b) near the PSB (EST) readout position for $Q_L$ (see Supplementary Note 3 for measurement result and fidelity analysis of $Q_R$). As expected, the lifetime $T_1$ of the $T_0$ state at the PSB region is in the order of 200 ns, indicating strong spin state mixing and subsequent charge tunneling due to the large $\Delta B_{//}$ produced by the micromagnet (see Supplementary Note 4 for magnetic field simulation). However, at large negative $\varepsilon$, the PSB is eventually lifted, and the absence of rapid spin mixing as well as the insensitivity of the $(2,0)T_0 - (2,0)S$ spin splitting to charge fluctuations ensures the long lifetime of the $T_0$ state. The evolution time at O is varied in the EST relaxation time measurement in Fig. 2b, and the amplitude decay of the coherent oscillation is probed to remove background signals typically present for long pulse repetition periods. The resultant $T_1$ of 337 $\mu s$ is three orders of magnitude longer than that in PSB readout. Without fast $\varepsilon$ modulation, a long $T_1$ exceeding 2.5 ms has been reported in GaAs QDs[43] implying that further optimization is possible.

**Measurement fidelity optimization**

Next, we discuss the calibration of the tunnel rates for single-shot readout and the optimization of the readout fidelity and visibility with the given experimental parameters. While for time-averaged charge detection we use a minimum integration time of 30 ns in the signal demodulation setup, corresponding to a measurement bandwidth of 33 MHz, we set the integration time to 1 $\mu s$ for single-shot detection to increase the signal to noise ratio, and we typically tune the tunneling rates to less than 1 MHz. Fig. 2c shows time-resolved tunnel out events triggered by the end of the pulse sequence from which we measure the tunneling out rate $\nu_{\text{out}} \sim \tau_{\text{out}}^{-1} = (16\mu\text{s})^{-1}$, extracted from the fit to an exponentially decaying function. The rate is within our measurement bandwidth. Also note that the ratio $T_1/\tau_{\text{out}}$ is at least 20, which is reasonable to perform high-fidelity measurements above 90% [44]. Fig. 2d shows the resultant histogram showing a separation of the mean value of the S and $T_0$ signal levels of more than 8 times the standard deviation, confirming the high fidelity of single-shot spin state detection with 1 $\mu$s integration time. We also find good agreement between the experimental and numerically simulated single-shot histograms[3] generated using the measured tunneling rates and signal to noise ratio (See Supplementary Note 5 for details).

After the rf demodulation stage, we further apply correlated double sampling (CDS)[15] to the single-shot traces to simplify the state discrimination and measurement automation. Using a fast boxcar integration with two gate windows that are 5 $\mu s$ apart in the time domain, a dc background-removed pseudo-time derivative of the single-shot traces is generated, enabling separate detection of tunneling out/in events with an external pulse counter (Stanford Research Systems, SR400 dual gated photon counter) and time-correlated pulse counting with a multichannel scaler (Stanford Research Systems, SR430 multichannel scaler) without the need for customized field-programmable gate array (FPGA) programming[37,45] (see Supplementary Note 2 for details of the CDS scheme). While this scheme was successful, the electronic measurement bandwidth was further reduced to 200 kHz for single-shot detection, which resulted in a relatively long readout time requiring relatively slow tunneling rates. To simulate realistic measurement conditions, we applied the numerical CDS filter to the simulated single-shot traces (Fig. 2e) and reproduced the tunneling detection fidelity of the measurement setup. As the measured electron temperature (230 mK ~ 5~6 GHz h/$k_B$, where h is the Planck's constant and $k_B$ is the Boltzmann's constant) compared to the ST$_0$ splitting (25~30 GHz) may trigger unwanted events such as false initialization, thermal tunneling of the ground state, and double-tunneling events within the measurement windows, we have introduced corresponding thermal parameters to the analysis. The parameters were utilized to model the Larmor oscillation measured (see Fig. 3), and the values were extracted from the least squares fitting with the experimental data to yield the final measurement fidelity (see Supplementary Note 5 for measurement fidelity analysis). The resulting theoretical measurement fidelity of the $Q_L$ is $90 \pm 1.3$ %, corresponding to a visibility of $80 \pm 2.6$ %, confirming that high-fidelity single-shot detection is possible at the given experimental conditions. Moreover, in Supplementary Note 6, we show through numerical simulation that FPGA-based single-shot detection, which we plan to perform in the future, will yield a measurement fidelity (visibility) of 94% (89%) at the same experimental condition through faster and more accurate peak detection which lowers the tunneling detection infidelity.

**High-visibility quantum control with the EST readout**

We now demonstrate high-visibility coherent qubit operations with the EST single-shot readout. The panels in Fig. 3 show the high-visibility two-axis control of $Q_L$ (Figs. 3a–c) and $Q_R$ (Figs. 3d–f) under large $\Delta B_{//}$ recorded with a single rf-set. For the $\Delta B_{//}$ oscillations (Figs. 3a, 3d), the I – W – O – W – R with the period of 150 $\mu s$ (Fig. 3a, top panel) was applied, and the evolution time at O was varied from 0 to 10 ns. Each trace in Figs. 3a and 3d is the average of 50 repeated measurements with 2000 shots per point, which takes over 5 min; thus, we expect an ensemble-averaged coherence time of $ST_0$ qubit oscillation $T_2{}^*$ in the order of 15 ns, limited by nuclear bath fluctuation[1]. We clearly observe coherent oscillations of $Q_L$ ($Q_R$) with ~81% (~64%) visibility, which is consistent with the results of the numerical simulation reported in Supplementary Note 5. Under the large $\Delta B_{//}$ of 100 (80) mT, corresponding to an oscillation frequency of 500 (400) MHz, we expect the Q-factor ($T_2^*/T_\pi$) of the oscillation to reach up to 28 (22) for $Q_L$ ($Q_R$), even with the measured ensemble-averaged $T_2^* \sim 15$ ns. Moreover, we estimate the leakage probability during the fast ramp less than 2% (see Supplementary Note 7); thus, we assume here that the effect of leakage error to the visibility is not significant. As discussed above, electronic bandwidth owing to the CDS technique is one of the factors limiting the visibility for both $Q_L$ and $Q_R$. Moreover, we estimate about 8% (9%) probability that the ground state tunnel out to the reservoir and 4% (2%) probability of false initialization to $T_0$ state for $Q_L$ ($Q_R$), showing that the reduction of the measurement fidelity and visibility in our experiment stems from the combination of the thermal effects, spin relaxation, and electronic bandwidth of the CDS method. For $Q_R$, tuning to an even longer $\tau_{out}$ of 25 $\mu s$ was necessary to account for the reduced rf-set sensor's signal contrast to farther QDs, for which the final visibility is approximately 64%. However, as shown in Supplementary Note 6, the visibility of the further QDs can be easily enhanced to more than 78% by simply improving the electronics of the measurement system, for example, with FPGA programming.

To acquire the 2D plots shown in Figs. 3b and 3e, the typical Ramsey pulse sequence of $I - W - O\,(\pi/2) - A_{ex} - O\,(\pi/2) - W - R$ (Fig. 3b, top panel) was applied, and the detuning amplitude $A_{ex}$ and evolution time $\tau_{ex}$ at the exchange step were varied. The figures show high-visibility quantum oscillation as well as continuous evolution of rotation axis on the Bloch sphere as $A_{ex}$ is varied over different regimes, where $T_2^*$ is limited by the charge noise for $J(\varepsilon) > \Delta B_{//}$ or by fluctuations in $\Delta B_{//}$ for $J(\varepsilon) \sim 0$. The fast Fourier transform (FFT) of the exchange oscillations along the exchange detuning axis (Figs. 3c and 3f) confirms the control of the $ST_0$ qubit over the two axes on the Bloch sphere for both $Q_L$ and $Q_R$, which is consistent with the expected qubit energy splitting (Fig. 3c, top panel). We emphasize that the measurement of two qubits is possible with one accompanied rf-set, which can be useful for the linear extension of the $ST_0$ qubits because the charge sensor does not need to be aligned with the QD array. In this work, we focused on independent two single-qubit gate operation; nevertheless, we expect that long $T_1$ at EST readout will allow the sequential measurement of two qubit states for a given quantum operation, which, in turn, will allow two qubit correlation measurement, enabling full two qubit state and process tomography in the future. Characterization of the two qubit interaction of $ST_0$ qubits in the current quadruple dot array, for example by dipole coupling[6,10] or exchange interaction[36], is the subject of current investigations.

**Discussion**

High-visibility readout of the $ST_0$ qubit at large $\Delta B_{//}$ is necessary for high-fidelity $ST_0$ qubit operations[6,37]. We performed high-visibility single-shot readout of two adjacent $ST_0$ qubits at $\Delta B_{//}$ of 100 mT (~$500\,MHz \cdot h \cdot (g * \cdot \mu_\mathrm{B})^{-1}$) by direct EST with one rf-set. No mixing between $T_0$ and (1,1)S state was observed at the EST readout point, which would allow sequential readout of multiple arrays of qubits due to the long $T_1$. Full one-electron signal difference discriminates the S and $T_0$ states compared to other readout methods where the dipolar charge difference is measured to readout the $ST_0$ qubit states[13,16]. This feature can be especially useful for scaling up the $ST_0$ qubits for the following reasons: 1) the large signal contrast can result in high visibility and low measurement error, and 2) the sensor does not need to be aligned along the QD array. Especially for GaAs spin qubits, high-visibility $ST_0$ qubit readout allows fast nuclear-spin fluctuation measurements, which will enable accurate feedback/stabilization of the nuclear spin bath for high-fidelity qubit control[2,32,37]. Furthermore, our method does not require additional metastable states[15,17,46] or pulsing sequences for high-fidelity measurements at large $\Delta B_{//}$ [14,16], showing that the experimental complexity is greatly reduced. EST readout of $ST_0$ qubits in nuclear spin-free systems, including Si, may also enhance the measurement fidelity by providing even longer $T_1$ for electron spins[7,47,48]. We further expect that the large $\Delta B_{//}$ based high-fidelity control combined with the high-fidelity readout method will be a powerful tool not only for single-qubit operations but also for exploring the charge-noise insensitive two-qubit operations of the $ST_0$ qubits using extended sweet spot[6].

Because the highest bandwidth potential of rf-reflectometry cannot be fully exploited with the CDS technique used in this study, we expect that the use of FPGA to detect the peaks from the bare rf demodulated single-shot traces will enhance the visibility to at least 88% (78%) for $Q_L$ ($Q_R$). The use of FPGA programming will also allow faster nuclear environment Hamiltonian learning[37], which can be useful in, for example, studying the time-correlation of nuclear spin bath fluctuations at different QD sites. We have taken the thermal tunneling probabilities into the analysis, and have successfully modeled the coherent $ST_0$ oscillation in our measurement setup, and derived the measurement fidelities. In the future, we plan to improve the performance by adopting an FPGA-based customized measurement, reducing electron temperature, and further optimizing the electronic signal path. However, even with the current limitations, the achieved visibility of 81% for $ST_0$ qubits at large $\Delta B_{//}$ shows potential to realize high-fidelity quantum measurements in scalable and individually addressable multiple QD arrays in semiconductors.

## Methods

**Device Fabrication.** The quadruple QD device was fabricated on a GaAs/AlGaAs heterostructure with a 2DEG formed 73 nm below the surface. The transport property of the 2DEG shows mobility $\mu = 2.6 \times 10^6 cm^2 V^{-1} s^{-1}$ with electron density $n = 4.6 \times 10^{11} cm^{-2}$ and temperature $T$ = 4 K. Mesa was defined by the wet etching technique to eliminate the 2DEG outside the region of interest. Ohmic contact was formed through metal diffusion to connect the 2DEG with the electrode on the surface. The depletion gates were fabricated on the surface using standard e-beam lithography and metal evaporation. The QD array axis was oriented parallel to the [011] crystallographic direction of GaAs. Subsequently, the micromagnet was patterned perpendicular to the QD array using standard e-beam lithography, and a Ni 10 nm/Co 250 nm/Au 5 nm was deposited using metal evaporation.

**Measurement.** The experiments were performed on a quadruple QD device placed on the 20 mK plate in a commercial dilution refrigerator (Oxford instruments, Triton-500). Rapid voltage pulses generated by Agilent M8195A arbitrary waveform generator (65 GSa/s sampling rate) and stable dc voltages generated by battery-operated voltage sources (Stanford Research Systems SIM928) were applied through bias-tees (picosecond Pulselabs 5546) in the dilution refrigerator before applying the metal gates. An LC-resonant tank circuit was attached to one of the ohmic contacts near the rf-set with a resonance frequency of ~110 MHz for homodyne detection. The reflected rf-signal was first amplified at 4 K with a commercial cryogenic amplifier (Caltech Microwave Research, CITLF2) and then further amplified at room temperature with home-made low-noise amplifiers. Signal demodulation was performed with an ultra-high-frequency lock-in amplifier (Zurich instrument UHFLI), and the demodulated amplitude was processed using a boxcar integrator built in the UHFLI for CDS. The CDS peaks were counted with an external photon counter (Stanford Research, SR400). The pulse parameters could be rapidly swept via a hardware looping technique, which enabled fast acquisition of the $\Delta B_{//}$ oscillations. In Supplementary Note 8, we show the details of the measurement setup, CDS technique, and signal analysis.

**Data Availability**

The data that support the findings of this study are available from the corresponding author upon request.

**Acknowledgements**

This work was supported by Samsung Science and Technology Foundation under Project Number SSTF-BA1502-03. We thank the Mark Eriksson group for their technical support on the circuit board used in the experiment.

**Competing interests**

The authors declare no competing interests.

**Author contributions**

D.K. and W.J. conceived the project, performed the measurements, and analyzed the data with the help of S.P. J.K. fabricated the device with the help of H.C., J.E, and Y.C. M.C. and W.J. built the experimental setup and configured the measurement software. V.U. synthesized and provided the GaAs heterostructure. All authors contributed to the preparation of the manuscript.

**Materials & Correspondence**

Correspondence and requests for materials should be addressed to D.K. ([dohunkim@snu.ac.kr](mailto:dohunkim@snu.ac.kr)).

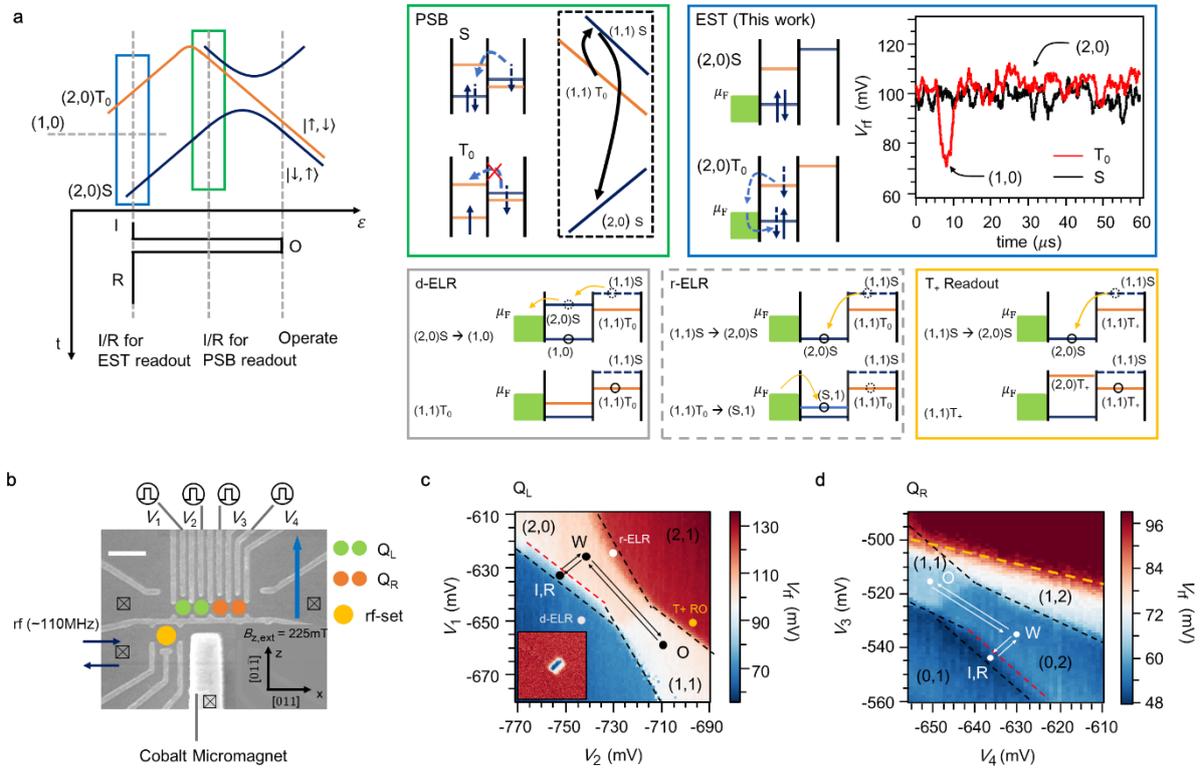

**Figure 1. Energy levels and device platform a.** Schematic of the singlet–triplet (ST$_0$) qubit energy levels as a function of detuning $\varepsilon$ with energy selective tunneling (EST, blue boxes)- and Pauli spin blockade (PSB, green boxes)-based readout schemes. Green panel: At the PSB readout point, the (1,1)S state tunnels into the (2,0) charge state while the tunneling from the (1,1)T$_0$ state is blocked. The relative charge position is observed to determine the qubit state. Finite magnetic field difference $\Delta B_{//}$ provides a relaxation pathway for the (1,1)T$_0$ state. Blue panel: The Fermi level resides between the (2,0)S and (2,0)T$_0$ states, which enables EST. The triplet state (red) tunnels out to the (1,0) and initializes to the (2,0)S, while no tunneling occurs for the S state. For comparison, three other types of modified PSB readout scheme are presented. Grey panel: In the direct enhanced latched readout (d-ELR) scheme, the (2,0)S state tunnels out to (1,0) while the spin-blockaded (1,1)T$_0$ state cannot tunnel out[14]. Dashed grey panel: At the reverse ELR (r-ELR) point, an electron tunnels into the spin-blockaded $T_0$ state to form the (S,1) while the S state stays at the (2,0) [14,15,17]. Yellow panel: In the $T_+$ readout scheme [16], one of the qubit states is conversed into the $T_+$ state to prevent the relaxation in the PSB. The readout is taken in the PSB by discriminating the (2,0)S and (1,1)$T_+$. Points corresponding to different schemes are denoted in Fig. 1c. **b.** Scanning electron microscopy image of the device. Green (orange) dots indicate the left (right) $ST_0$ qubit $Q_L$ ($Q_R$), and the yellow dot indicates the rf single-electron transistor (rf-set). The blue arrow indicates the external magnetic field direction. The white scale bar corresponds to 500nm. **c.**(**d.**) Charge stability diagram for $Q_L$ ($Q_R$) operation with the pulse cycling I – W – O – W – R points superimposed. The red dashed line shows the boundary of the region inside which the EST readout is appropriate. The inset of **c.** shows the PSB readout signal for the same area observed by gated integration. The yellow line in **d.** shows the electron transition signal of the QD coupled to $V_2$.

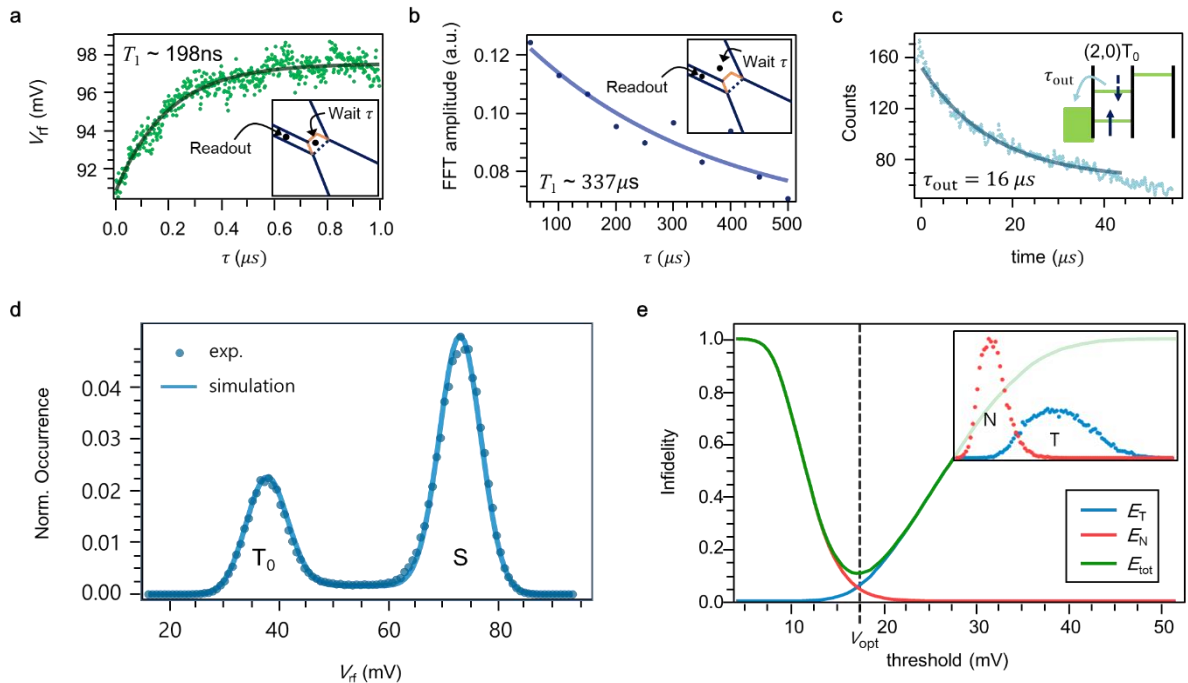

**Figure 2. Time-resolved relaxation measurements and fidelity analysis of $Q_L$ a.** Relaxation time measurement at PSB readout. The time-averaged rf-demodulated signal $V_{rf}$ is recorded as a function of the waiting time $\tau$ at the $\varepsilon$ denoted in the inset. $T_1 \sim 200$ ns is extracted from the fitting data to the exponential decay curve. **b.** Relaxation time measurement near EST readout. The decay of the coherent oscillation is observed along the waiting time $\tau$ near the detuning at the measurement point denoted in the inset. $T_1 \sim 337\ \mu s$ is extracted. **c.** Histogram of the tunneling out events triggered by the end of the manipulation pulse as a function of time. **d.** Histogram of the experimental and simulated rf-demodulated single-shot traces with the application of $\pi$ pulses for EST readout showing a mean value separation of more than 8 times the standard deviation. **e.** Tunneling detection infidelity calculated from the CDS peak amplitude histogram shown in the inset. Minimum total error $(E_T + E_N)$ of ~10.5% corresponding to $E_T \sim 5\ \%$ , and $E_N \sim 5.5\ \%$ are estimated at the optimal threshold voltage $V_{opt}$.

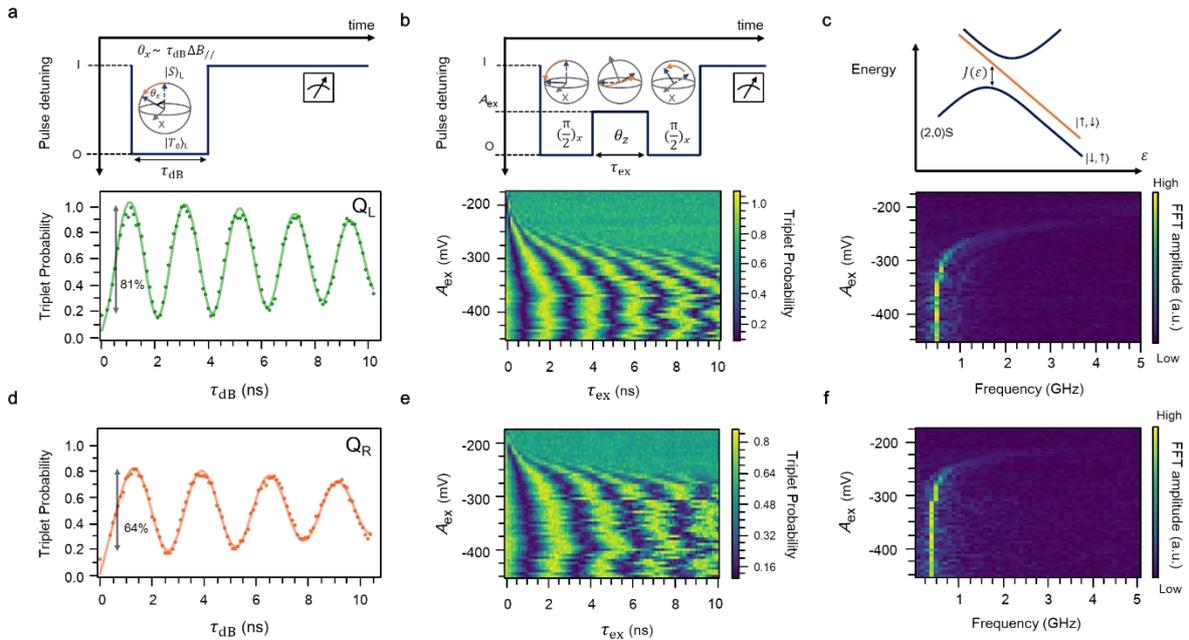

**Figure 3. High-visibility two-axis control of two ST₀ qubits. a.** (**d.**) Coherent ST$_0$

oscillation of Q$_L$ (Q$_R$) under large $\Delta B_{//}$. 81% (64%) quantum oscillation visibility is defined

by the initial oscillation amplitude which is in good agreement with the analytic model with

thermal effects and spin relaxation (See Supplementary Note 5). **b.** (**e.**) Coherent exchange

oscillation and two-axis control of Q$_L$ (Q$_R$) on the Bloch sphere. The top panel of **b.** shows

the Ramsey pulse sequence where the first $\pi/2$ pulse induces equal superposition of S and

T$_0$ spin states, and the phase evolution under non-zero $J(\varepsilon)$ is probed by the second $\pi/2$

pulse. By varying the pulse amplitude $A_{\mathrm{ex}}$ and the evolution time $\tau_{\mathrm{ex}}$ at the exchange step,

the high-resolution rotation axis evolution and an energy spectrum consistent with the

expected functional form of $J(\varepsilon)$ [38], the schematic of which is shown in the top panel of

**c.**, are confirmed by the fast Fourier transform (FFT) plots in **c.** (**f.**).

# Robust energy selective tunneling readout of singlet triplet qubits under large magnetic field gradient

Wonjin Jang[1†], Jehyun Kim[1†], Min-Kyun Cho[1], Hwanchul Chung[2], Sanghyeok Park[1], Jaeun Eom[1], Vladimir Umansky[3], Yunchul Chung[2], and Dohun Kim[1]*

[1]Department of Physics and Astronomy, and Institute of Applied Physics, Seoul National University, Seoul 08826, Korea

[2] Department of Physics, Pusan National University, Busan 46241, Korea

[3]Braun Center for Submicron Research, Department of Condensed Matter Physics, Weizmann Institute of Science, Rehovot 76100, Israel

[†]These authors contributed equally to this work

*Corresponding author: dohunkim@snu.ac.kr

## Supplementary Information

### Supplementary Note 1. Electron temperature and intra-qubit tunnel coupling calibration

Electron temperature, and the tunnel coupling strength of the left double quantum dot are measured using the standard lock-in technique. $dV_{rf}/dV_2$ is observed by modulating $V_2$ gate voltage with 337Hz frequency. With proper adjustment of dot-reservoir tunnel rates less than 1 MHz and setting minimal modulation amplitude, the electron temperature $T_e \sim 230$mK is determined by fitting the heterodyne detected single electron transition line to the equation

$$\frac{dV_{rf}}{dV_2}(V_1) = A_{offset} - \frac{A\alpha}{k_B T} \frac{\exp(\alpha(V_1 - V_{offset})/k_B T)}{(1 + \exp(\alpha(V_1 - V_{offset})/k_B T))^2}$$, which is the derivative of the typical

Fermi-Dirac distribution (Supplementary Fig. 1a). Here $\alpha = 0.035$ is the lever-arm of the $V_1$ gate obtained from the Coulomb diamond measurement, $k_B$ is the Boltzmann constant, and $A_{offset}$ and $V_{offset}$ are the $dV_{rf}/dV_2$ offset and the offset $V_1$ voltage in the $dV_{rf}/dV_2 - V_1$ plot, respectively. The intra-qubit tunnel coupling strength $t_c$ was obtained in the similar manner, by sweeping the gate voltage through the inter-dot transition line in the stability diagram for example shown in Fig. 1c of the main text. The broadening is fitted using the same equation described above, with the broadening width $2t_c$ instead of $k_B T$ where the $t_c$ represents the tunnel coupling strength. The resultant $2t_c/h$ is 16 GHz where $h$ is the Plank's constant.

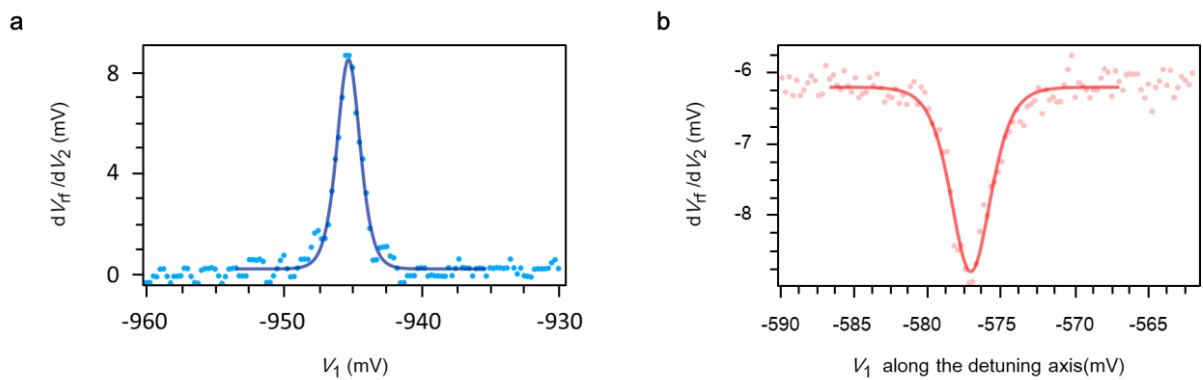

**Supplementary Figure 1. System parameter calibration. a.** Electron temperature measurement. **b.** tunnel coupling strength measurement using the heterodyne detection scheme. Typical lock-in measurement was performed to obtain the broadening of the single electron transition due to thermal broadening and the intra-qubit tunneling. Electron temperature $T_e \sim$ 230 mK, and tunnel coupling $t_c/h \sim$ 8GHz were obtained from the fitting. When obtaining **b.** both $V_1$, and $V_2$ were swept through the inter-dot transition line in Fig. 1c, but only the $V_1$ gate voltage is shown in the x-axis.

**Supplementary Note 2. Correlated double sampling (CDS)**

By resampling the demodulated rf-signal with the boxcar integrator, we enable the real-time single-shot event counting without the use of field-programmable gate arrays (FPGA) programming. As shown in Supplementary Fig. 2, the boxcar integrator subtracts the 100 ns-averaged baseline signal from the gate signal which are separated by 5 $\mu s$ in the time domain to yield a pseudo-time derivative signal of the single-shot trace with 200 kHz sampling rate. CDS converts the falling (rising) edge to the positive (negative) peak and the peaks are detected by the external photon counter (Stanford Research Systems SR400) as shown in Supplementary Fig. 2a. This allows the separate detection of tunneling in / out event in real-time without post-processing which may reduce the experimental overhead in the analysis step. By counting the tunneling out events, we have observed the coherent singlet-triplet qubit ($ST_0$ qubit) oscillations in the energy selective tunneling (EST) readout point in the main text. For single-shot readout, the boxcar integrator is operated with average number set to 1 (no averaging).

When averaged, however, the CDS technique can also be utilized to observe short-lived $T_0$ signal for Pauli Spin Blockade (PSB) readout, which enable measurement bandwidth of 33MHz in time averaged manner (see also the inset to Fig. 1c in the main text). By setting the $\sim 0.1$ $\mu s$ gate window right after the spin-mixing pulse comes back to the PSB region, and the $\sim 0.1$ $\mu s$ baseline gate window before the next pulse start as shown in Supplementary Fig. 2b, the demodulated signal is effectively sampled for short time where the portion of the $T_0$ signal is sufficiently large to be observed with sufficient periodic average.

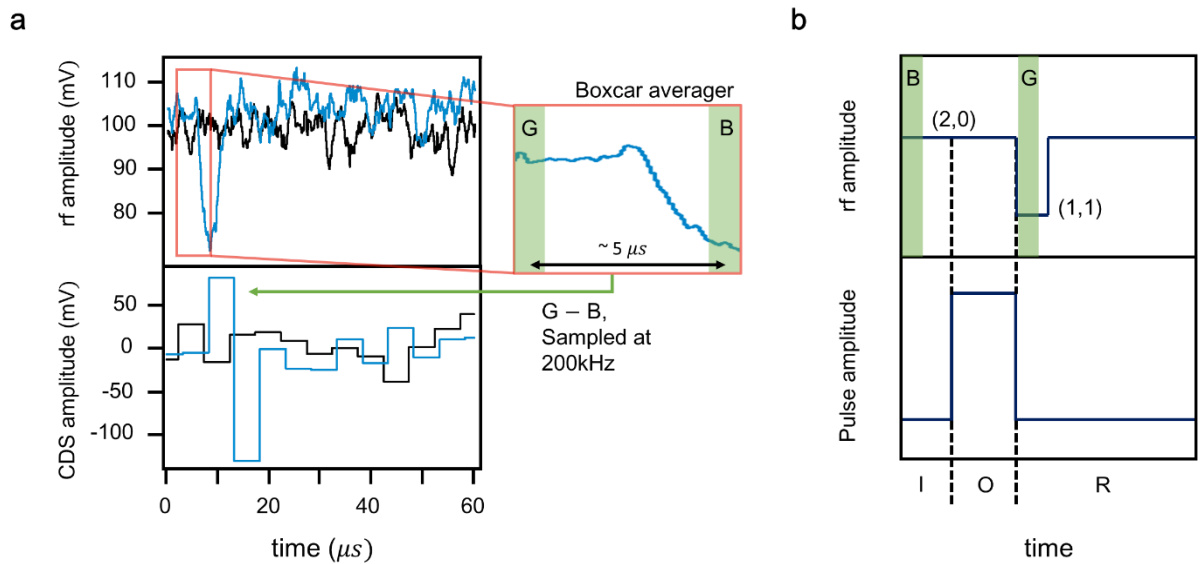

**Supplementary Figure 2. Correlated double sampling schematics. a.** Correlated double sampling for tunneling out / in event detection. Boxcar integrator resamples the bare demodulated rf signal by subtracting the ~ 100 ns averaged baseline (B) signal from the gate (G) signal every 5 $\mu s$ . This resampling process converts the falling edge signal of the rf signal to a positive peak with removing dc background and produces pulse signal robust to background drift. **b.** CDS scheme for short $T_0$ signal detection in PSB readout. Pulse mixes the S and $T_0$ states in the operation (O) sequence, and when returning to the readout (R) step, the $T_0$ quickly relaxes to (2,0) charge state under large magnetic field difference. The boxcar integrator in this case is operated in averaging mode where sampled signal G of the rf-signal for short period time after the pulse sequence are subtracted by the B signal and averaged about 5000 times to increase signal to noise ratio.

**Supplementary Note 3. Right qubit measurement fidelity**

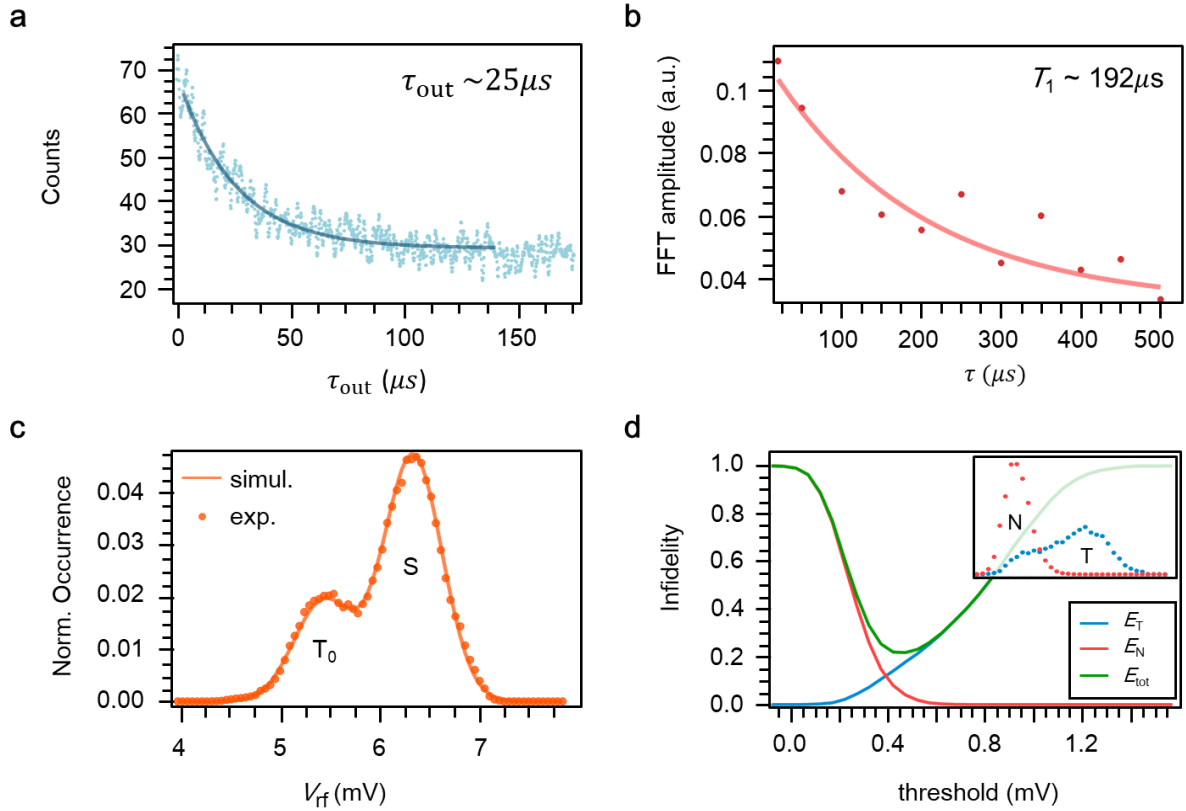

**Supplementary Figure 3. Right qubit readout fidelity analysis. a.** Tunneling out rate of the right qubit $Q_R$ at the EST readout point. Tunneling out events were recorded as a function of the tunneling time, and the exponential fit to the curve yields $\tau_{out} \sim 25\mu s$. **b.** Relaxation time measurement near EST readout point. The decay of the coherent oscillation is observed along the waiting time $\tau$ near the EST readout point. $T_1 \sim 192\,\mu s$ is extracted from the fit. **c.** Experimental, and simulated rf single-shot traces of the $Q_R$ with the $\pi$-pulse applied. **d.** Tunneling detection infidelity calculated from the CDS peak amplitude histogram shown in the inset. Minimum total error ($E_T + E_N$) of 28.2% corresponding to $E_T \sim 19\%$, and $E_N \sim 9.2\%$ are estimated at the optimal threshold.

**Supplementary Note 4. Magnetic field simulation**

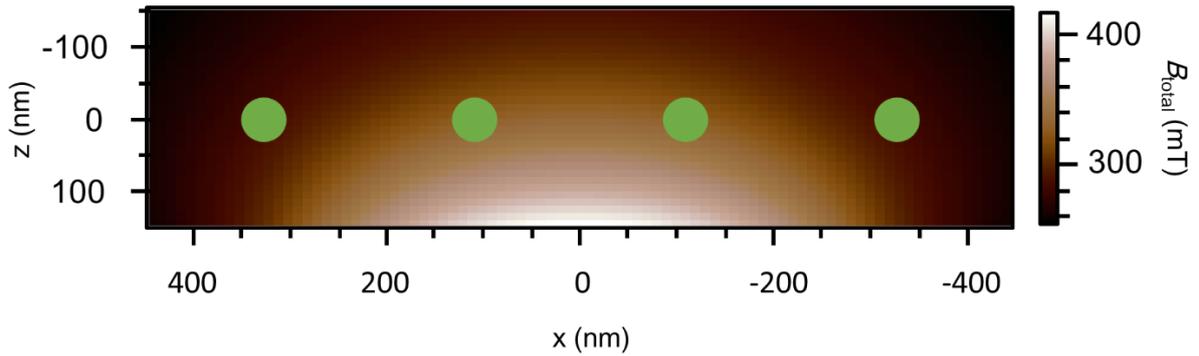

**Supplementary Figure 4. Simulation of the magnetic field around the QDs.** The total magnetic field strength around the quantum dots in our device (see Supplementary Fig. 8) is simulated using the boundary integral method with RADIA[1,2] package. Green dots indicate the quantum dot positions. The fast $\Delta B_{//}$ oscillations shown in Fig. 3 in the main text is up to 500MHz corresponding to $\Delta B_{//}$ of 100 mT, and we ascribe this higher-than-expected-$\Delta B_{//}$ to the displacement of the electrons from the expected positions by the confining potential in the few electron regime.

**Supplementary Note 5. Measurement fidelity analysis**

We have taken the thermal tunneling events into consideration for the fidelity analysis and describe the analysis protocol in detail here. We first define two parameters $\alpha_1$, and $\beta$ where $\alpha_1$ corresponds to the probability for the ground (S) state to tunnel out to the reservoir within a measurement window, and $\beta$ corresponds to the false initialization probability following the Pla. *et al.*[3]. Regarding the false initialization we assume the following for three triplet states – $T_0$, $T_+$, and $T_-$.

1) Probabilities for the electron to falsely initialize to different triplet states are all equal to $\beta/3$.

2) The relaxation time is equal for all $T_0$, $T_+$, and $T_-$ state.

3) $T_+$ (1,1), and $T_-$ (1,1) states do not evolve to other states during the Larmor oscillation phase.

It should be noted that while the false initialization to $T_0$ state contribute to the visibility loss while the false initialization to $T_+$ or $T_-$ states would result in overall shift of the Larmor oscillation because the $T_+$ or $T_-$ will not undergo coherent mixing process during the evolution time. We introduce an additional parameter, $\alpha_2$ to account for the double-tunneling probability of the ground state within a single measurement window. For example, in the case that a $T_0$ state first tunnels out to the reservoir and initialize to the S state in a measurement phase, there still exist non-zero probability for the S state to tunnel out within the measurement window, and $\alpha_2$ represents the corresponding probability. It is thus natural to define the total double tunneling probability as $P_2 = (\beta + (1-\beta)\alpha_2)$ which covers the double-tunneling probability of the false initialized triplet states and the reinitialized S state after a single tunneling event.

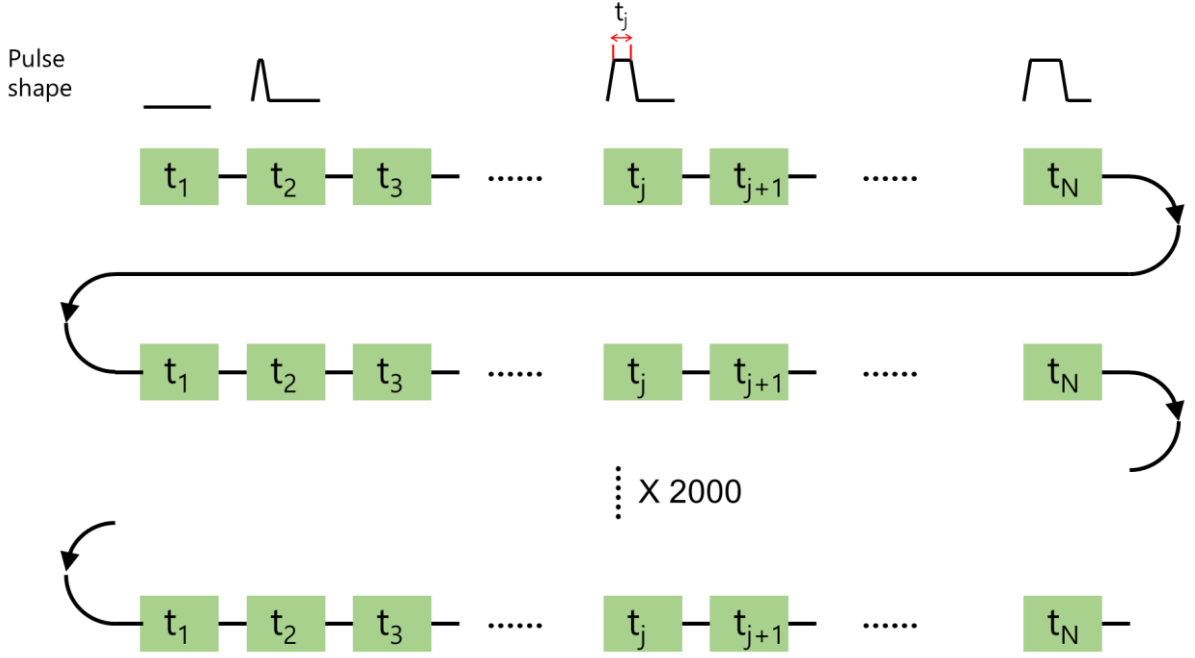

**Supplementary Figure 5. Pulse sequence for Larmor oscillation measurement.** The Larmor oscillations of $Q_L$ and $Q_R$ are measured by first sweeping the pulse parameter, the free evolution time $t_j$, and repeating the measurement over 2000 times to average traces. N different pulses corresponding to N different evolution time are all recorded in the arbitrary waveform generator (AWG) before measurements to enable the rapid hardware triggered sweep of the pulse parameter.

In the Larmor experiment in Supplementary Fig. 3a, 3d of the main text, we obtain the oscillation by averaging single-shot traces using the pulse sequence shown in Supplementary Fig. 5. As we regard the spin state is at the excited ($T_0$) state if there is at least one tunneling event within a measurement window, we first define the $P(t_j, \Delta B_{//})$ as the probability for at least one tunneling to occur within a single measurement window at the evolution time $t_j$ ($1 \le j \le N$, j is integer) under the magnetic field difference $\Delta B_{//}$. It should be noted that $P(t_j, \Delta B_{//})$ must be derived recursively since the tunneling event at the j$^{th}$ shot affects the tunneling probability of the (j+1)$^{th}$ shot. The relation between the $P(t_{j+1}, \Delta B_{//})$, and $P(t_j, \Delta B_{//})$ is as follows.

$$P(t_{j+1}, \Delta B_{//}) = P(t_j, \Delta B_{//})[(1-\beta)\{f_{j+1}r + (1-f_{j+1})\alpha_1 + f_{j+1}(1-r)\alpha_1\} +$$
$$\frac{\beta}{3}\{f_{j+1}\alpha_1 + (1-f_{j+1})r + (1-f_{j+1})(1-r)\alpha_1\} +$$

$$\frac{2\beta}{3}\{r+(1-r)\alpha_1\}]+$$

$$(1-P(t_j,\Delta B_{//}))(f_{j+1}r+(1-f_{j+1})\alpha_1+f_{j+1}(1-r)\alpha_1) \qquad -(1)$$

Here $f_{j+1} = f(t_{j+1}, \Delta B_{//}) = \sin^2(\pi\Delta B_{//}t_{j+1})$ is the ideal $T_0$ probability at the evolution time $t_{j+1}$ under the magnetic field difference $\Delta B_{//}$ when the initial state is the singlet state, and $(1-r)$ is the relaxation probability of the $T_0$ state within the measurement window which is given by

$r = \dfrac{\int_0^M \exp(-t/T_1)\exp(-t/\tau_{out})dt}{\int_0^\infty \exp(-t/\tau_{out})dt}$ where M is the length of the measurement window, $T_1$ is the spin-relaxation time, and $\tau_{out}$ is the tunneling-out time.

However, recursively obtained $P(t, \Delta B_{//})$ cannot yet fully account for the experimentally obtained Larmor curve. We additionally define tunneling detection fidelity $T_T$ ($T_N$) which is the fidelity to correctly tell there is a (no) tunneling event when there is a (no) peak in the signal. Here $T_T$ and $T_N$ are determined by the signal to noise ratio (SNR) of the measurement setup, and the detailed description on how to obtain the tunneling detection fidelities is given below. With $P(t, \Delta B_{//})$, $T_T$, and $T_N$, the experimental Larmor curve can be fully modeled. $A(t, \Delta B_{//})$, the average number of the tunneling events detected by the photon counter, has the following relation with the $P(t, \Delta B_{//})$.

$$A(t,\Delta B_{//}) = P(t,\Delta B_{//})(1+P_2)T_T + (1-P(t,\Delta B_{//}))(1-T_N) \qquad -(2)$$

Assuming that $\Delta B_{//}$ suffers from the Gaussian noise, we perform the Gaussian weighted sum of $A(t, \Delta B_{//})$ curves as below within the 5-sigma range.

$$\overline{A}(t,\Delta B_{//}) = \sum_{b=\Delta B_{//}-5\sigma}^{\Delta B_{//}+5\sigma} A(t,b)\,G(b,\Delta B_{//},\sigma)\Delta b \qquad -(3)$$

Here $G(x, \mu, \sigma)$ is the Gaussian distribution centered at $\mu$ with the standard deviation $\sigma$.

By setting $\alpha_1, \alpha_2, \beta, \sigma$, and $\Delta B_{//}$ as the fitting parameters we perform the least squares fitting of the $\overline{A}(t, \Delta B_{//})$ to the experimental Larmor curve. Below we describe the protocol for obtaining the tunneling detection infidelities.

Typical measurement fidelities are acquired by obtaining the histograms of the time-resolved signals of qubit ground and excited states, and finding the adequate threshold which yields the highest visibility[4–6]. The obtained measurement fidelities not only suffer from the imperfect tunneling detection, but also from the spin-relaxation or thermal tunneling events, implying that the $T_T$, and $T_N$ cannot be solely obtained experimentally. We first numerically simulate[6] the traces with the experimental parameters including the offset rf-voltage, amplitude of the tunneling peaks, tunneling in/out time, spin-relaxation ($T_1$) time, and sampling rate (Parameters are denoted in the Supplementary Table 1.). The thermal tunneling events are added to the signals in according to the thermal tunneling parameters $\alpha_1$, $\alpha_2$ and $\beta$, which then undergo through the numerical noise and low pass filter to yield a realistic signal. Then the amplitude of the noise filter is varied to match the experimentally obtained histogram of the rf-signal as in Fig. 2d, and the optimal noise amplitude is chosen. With the noise amplitude, we numerically generate the 'ideal' signals of triplets and singlets without the thermal tunneling events, or spin-relaxation to solely evaluate the tunneling detection fidelity of the electrical measurement setup. As we have utilized the CDS technique as described in Supplementary Note 2, corresponding boxcar filter is applied to the numerical signals, and the histograms of the boxcar-filtered signals are acquired to perform a typical integration for tunneling detection fidelity calculation[4–6]. We have plotted the tunneling detection infidelity $E_T$ ($E_N$) where $E_T = 1 - T_T$ ($E_N = 1 - T_N$) in the Fig.2e and Supplementary Fig. 3d. The tunneling detection fidelities $T_T(V_{op})$, and $T_N(V_{op})$ at the optimal threshold which yields the lowest $E_{tot}(V_{op}) = E_T(V_{op}) + E_N(V_{op})$ are utilized for the Larmor curve fitting described above.

To sum up, the whole process is done as follows.

1) Put the initial guesses of parameters to perform Larmor curve fitting, and obtain the $\alpha_1, \alpha_2$, and $\beta$
2) Use the obtained thermal tunneling parameters for rf-histogram fitting to acquire the optimal noise amplitude.
3) Generate ideal traces of the $T_0$, and S states with the noise amplitude from 2), and calculate $T_T$, and $T_N$
4) Use $T_T$, and $T_N$ for Larmor curve fitting, and obtain $\alpha_1, \alpha_2$, and $\beta$.
5) Iteratively obtain the optimal $T_T$, $T_N$, $\alpha_1, \alpha_2$, and $\beta$.

We now turn to discuss the total measurement fidelity. If there exist thermal tunneling events irrelevant with the spin dynamics, it is difficult to tell whether the tunneling peak occurs due to the thermal effect or not upon acquiring a single-shot trace. Thereby the total measurement fidelity should now be obtained by taking $\alpha_1, \alpha_2$, and $\beta$ into account. Let us define $F_{T_0}$ ($F_S$) as the $T_0$ (S) measurement fidelity, and $R_{T_0}(R_S) = 1 - F_{T_0}(F_S)$ as the measurement infidelity. We first evaluate $R_S$ by categorizing the cases which can detract the S measurement fidelity.

$X_1$: No tunneling occurs ($1 - \alpha_1$), photon counter 'beeps' due to electrical noise ($E_N$)

X$_2$: A single tunneling occurs ($\alpha_1$), photon counter detects the tunneling ($1-E_T$)

X$_3$: A single tunneling occurs ($\alpha_1$), the tunneling is not detected ($E_T$) but the photon counter 'beeps' due to electrical noise ($E_N$)

X$_4$: Double tunneling occurs ($\alpha_1 P_2$), first tunneling is not detected ($E_T$), and the second tunneling is detected ($1-E_T$)

X$_5$: Double tunneling occurs ($\alpha_1 P_2$), both tunneling events are not detected ($E_T{}^2$), but photon counter 'beeps' due to electrical noise ($E_N$)

As X$_1$ ~ X$_5$ are independent, mutually exclusive, $R_S = P(X_1) + P(X_2) + P(X_3) + P(X_4) + P(X_5)$ holds. i.e.

$$R_S = (1-\alpha_1)E_N + \alpha_1(1-E_T) + \alpha_1 E_T E_N + \alpha_1 P_2 E_T(1-E_T) + \alpha_1 P_2 E_T{}^2 E_N \qquad \text{- (4)}$$

Cases for the T$_0$ measurement infidelity can be similarly categorized with the relaxation process considered, as follows.

Y: T$_0$ relaxes within the measurement time ($1-r$), photon counter detects no tunneling ($1-R_S$)

Z$_1$: T$_0$ does not relax within the measurement time ($r$), the tunneling is not detected ($E_T$), no additional tunneling occurs ($1-P_2$), counter detects no signal ($1-E_N$)

Z$_2$: T$_0$ does not relax within the measurement time ($r$), double-tunneling occurs ($P_2$), both tunneling events are not detected ($E_T{}^2$)

Y, Z$_1$, Z$_2$ are all independent, and mutually exclusive leading to $R_{T_0} = P(Y) + P(Z_1) + P(Z_2)$. i.e.

$$R_{T_0} = (1-r)(1-R_S) + rE_T(1-P_2)(1-E_N) + rE_T{}^2 P_2 \qquad \text{- (5)}$$

Finally, the total measurement fidelity $F_{meas} = 1 - \dfrac{(R_S + R_{T_0})}{2}$ with the spin-relaxation, thermal tunneling events, and the tunneling detection infidelity of the setup is calculated as 90$\pm$1.3% (80.3$\pm$1 %) corresponding to visibility ($F_S + F_{T_0} - 1$) of 80$\pm$2.6 % (60.6$\pm$2 %) for Q$_L$ (Q$_R$). Also, from the Larmor curve fitting we obtain the $\Delta B_{//}$ fluctuation of $\sigma$ ~15.71 MHz (15.73 MHz) corresponding to $T_2^*$ ~ 14.33 ns (14.31 ns) for Q$_L$ (Q$_R$). We assume that ~ 3% disagreement of the Q$_R$ visibility is due to the uncertainty in measured relaxation time.

| Input | $Q_L$ | $Q_R$ |
|---|---|---|
| $\tau_{out}$ ($\mu s$) –:Tunneling-out time of the triplet states | 16 | 25.5 |
| $\tau_{in}$ ($\mu s$) : Tunneling-in time of the singlet state | 117 | 130.5 |
| $T_1$ ($\mu s$) : Relaxation time of the triplet states | 337 | 192 |
| Meas. Time ($\mu s$) | 150 | 200 |
| Sampling rate (MHz) | 14 | 14 |
| CDS freq. (kHz) | 200 | 50 |
| CDS gate width ($\mu s$) | 0.1 | 4 |
| **Output** | | |
| $\alpha_1$ : False tunneling-out probability of the singlet state | 0.081 | 0.092 |
| $\alpha_2$ : Double tunneling-out probability | 0.08 | 0.089 |
| $\beta$ : False initialization probability | 0.12 | 0.069 |
| $\sigma$ (MHz) : Std. deviation of the $\Delta B_{//}$ distribution | 15.71 | 15.73 |
| $E_T$ : Tunneling detection infidelity | 0.05 | 0.19 |
| $E_N$ : No-tunneling detection infidelity | 0.055 | 0.092 |
| $R_{T_0}$ : $T_0$ measurement infidelity | 0.077 | 0.232 |
| $R_S$ : S measurement infidelity | 0.128 | 0.162 |
| $F_{meas}$ : Total measurement fidelity | 90$\pm$1.3% | 80.3$\pm$1 % |

**Supplementary Table 1. Input and output parameters of the analysis**

**Supplementary Note 6. Expected fidelity with direct peak detection**

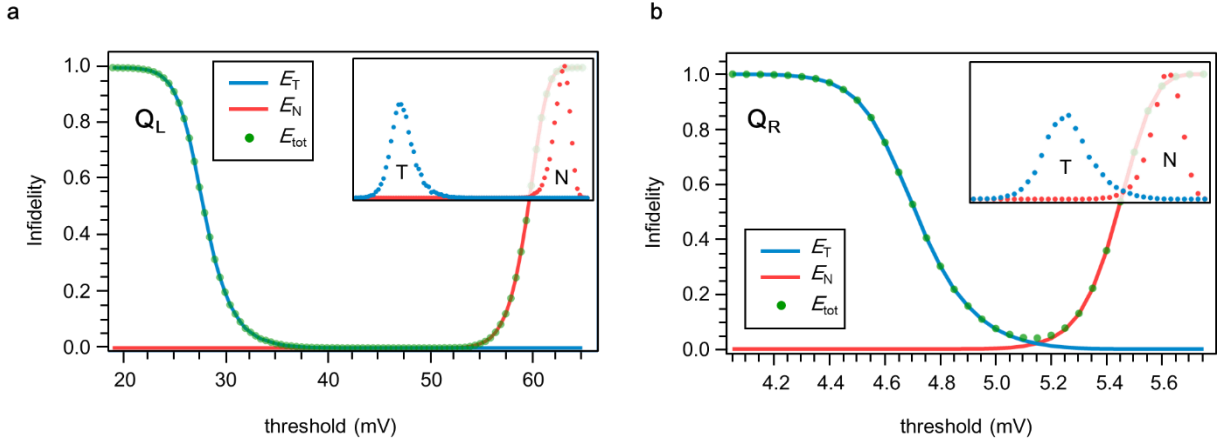

**Supplementary Figure 6. Error simulation for direct peak detection scheme. a.** (**b.**) The tunneling detection infidelity calculated from the rf-histogram in the inset. The histograms are constructed by sampling the peak values for $Q_L$ ($Q_R$) single-shot traces without the spin relaxation, and thermal tunneling events to evaluate the tunneling detection infidelities without the CDS. For $Q_L$, the tunneling detection infidelities are below 0.00001% while for $Q_R$ infidelities of $E_T \sim 2\%$ , and $E_N \sim 2\%$ are obtained at the optimal threshold.

The measurement fidelity and visibility are calculated for the direct peak detection scheme to explicitly show that the use of FPGA rather than CDS technique may extend the measurement fidelity and visibility with the same experimental parameters. Following the A. Morello *et al.*[6], single-shot traces were first simulated with the experimental parameters, and instead of passing through additional numerical CDS filter, the peak value (the minimum value) from each rf single-shot trace is sampled from 15,000 traces to construct the histogram shown in the insets of Supplementary Fig. 6a. and 6b. Because the short peaks or the full signal contrast cannot be perfectly detected with the CDS due to its limited bandwidth, the tunneling detection fidelities are naturally higher for the FPGA case. With the same $\tau_{out}$, $T_1$, $\alpha_1$ , $\alpha_2$ , and $\beta$ , the measurement fidelity of $Q_L$ ($Q_R$) is estimated as 94 % (88.8 %). We claim that the fidelities can further be higher if the FPGA-based readout is applied because the large peak separation would allow faster single-shot measurements with faster tunneling rates which would result in less relaxation due to lower $\tau_{out}/T_1$.

**Supplementary Note 7. Leakage error analysis due to Landau-Zener transition**

We estimate the Landau-Zener transition probability during the fast ramp time by solving the time-dependent Schrodinger equation with the typical $ST_0$ qubit Hamiltonian[7]. We put the measured parameters such as the tunnel coupling strength, pulse rise time, pulse amplitude, and the magnetic field differences into the numerical simulation, and obtained the time trace of (2,0)S along the evolution time up to 10 ns. As the decoherence of the system is not considered in the simulation, the resultant trace (Supplementary Fig. 7) exhibits non-decaying oscillatory behavior in the 0 ~ 3% range which averages to 1.7%. We therefore conclude that the leakage probability and its effect to the visibility is not significant.

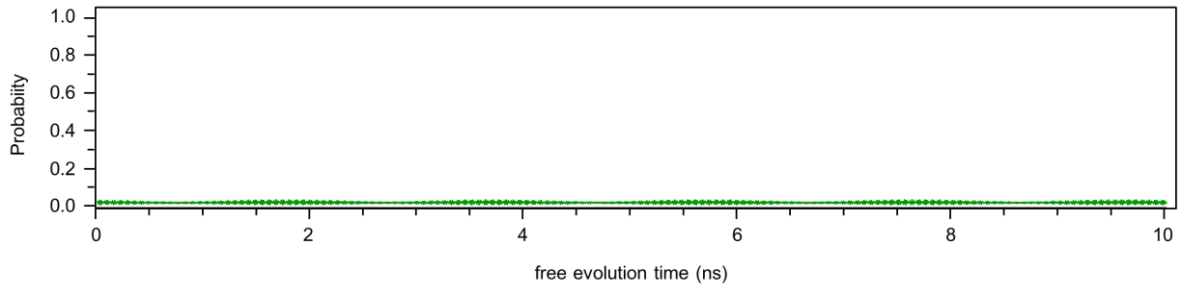

**Supplementary Figure 7. The (2,0)S probability along the free evolution time.** Time evolution of the (2,0)S state probability under the typical $ST_0$ qubit Hamiltonian is numerically obtained by putting the experimental parameters. The simulation yields 1.7% (2,0)S average occupation probability during the qubit manipulation time.

**Supplementary Note 8. Measurement setup**

A rf-single electron transistor (rf-set) sensor is operated to detect the charge states of the $ST_0$ qubits in our device. For the rf-reflectometry, impedance matching tank circuit as shown in Supplementary Fig. 8 is attached to the rf-ohmic contact of the device, and the 100 pF capacitor is connected in series to the other ohmic contact (depicted on the micromagnet) to serve as a rf-ground. With the inductor value L = 1500 nH and the parasitic capacitance $C_p$ = 1.4 pF of the circuit board, the resonance frequency is about 110MHz, and the impedance matching occurs at rf-set sensor resistance approximately 0.5 $h/e^2$ where $h$ is Plank's constant and $e$ is the electron charge. A commercial high frequency lock-in amplifier (Zurich Instrument, UHFLI) is used as the carrier generator, rf demodulator for the homodyne detection, and further signal processing such as gated integration and timing marker generation. Carrier power of -40dBm power is generated at room temperature and attenuated through the attenuators and the directional coupler by -50 dB in the input line. The reflected signal is first amplified by 25 dB with commercial cryogenic amplifier (Caltech Microwave Research Group, CITLF2), and further amplified by 50 dB at room temperature using a home-made low-noise rf amplifier. Demodulated signal is acquired with a data acquisition card (National Instruments, NI USB-9215A) for raster scanning and also boxcar-averaged with the gated integrator module in the UHFLI for the correlated double sampling described above. For single-shot readout, the CDS output is counted with a high-speed commercial photon counter (Stanford Research Systems, SR400 dual gated photon counter). A commercial multichannel scalar (Stanford Research Systems, SR430 multichannel scaler & average) is also used for time correlated pulse counting for tunneling rate calibration.

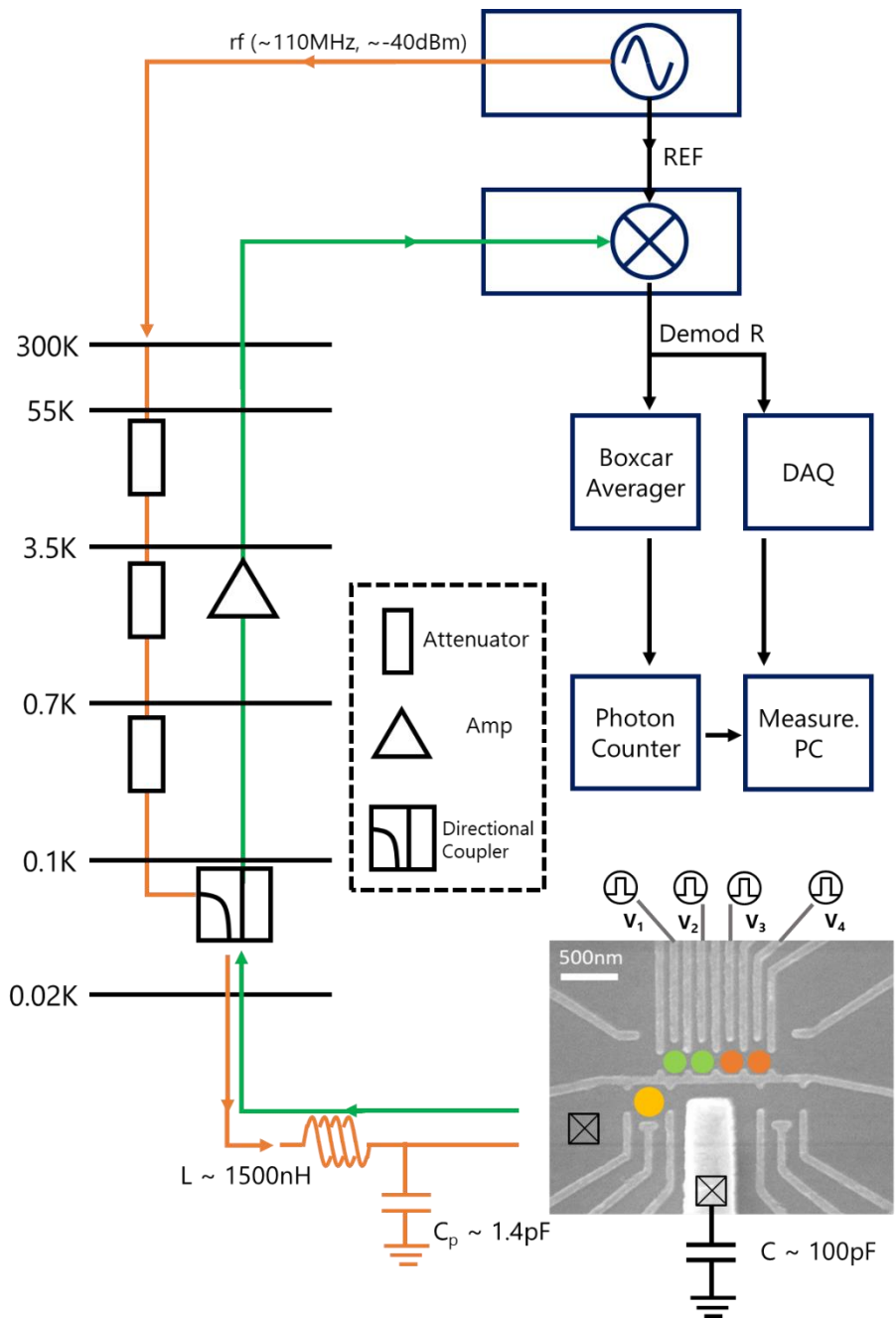

**Supplementary Figure 8. The measurement setup for radio frequency (rf)-reflectometry, and the signal block diagram.** Impedance matching tank-circuit ($L$~1500 nH, $C_p$ ~ 1.4pF) is attached to the rf-set sensor Ohmic contact for homodyne detection. Orange (green) line indicates the input (reflected) signal. Reflected signal is demodulated and processed for single-shot event counting as shown in the block diagram.

**Supplementary References**